\documentclass[reprint,showpacs,amsmath,amssymb,xcolor=dvipsnames,prl,superscriptaddress,floatfix]{revtex4-1}
\pdfoutput=1

\usepackage[T1]{fontenc}
\usepackage[pdftex]{graphicx}
\usepackage{bm}
\usepackage[pdftex,colorlinks=true,citecolor=blue,urlcolor=blue,linkcolor=black]{hyperref}
\usepackage{braket}
\usepackage[version-1-compatibility,binary-units,input-product=x,abbreviations=true]{siunitx}
\usepackage{multibib}

\newcommand*\diff{\mathop{}\!\mathrm{d}}

\makeatletter
\def\blfootnote{\gdef\@thefnmark{}\@footnotetext}
\makeatother

\begin{document}

%%%%%%%%%%%%%%%%%%%%%%%%%%%%%%%%%%%%%%%%%%%%%%%%%%%%%%%%%%%%%%%%%%%%%%%%%%%%%%%%%
%						    	Title and Authors	    						%
%%%%%%%%%%%%%%%%%%%%%%%%%%%%%%%%%%%%%%%%%%%%%%%%%%%%%%%%%%%%%%%%%%%%%%%%%%%%%%%%%

\title{Observation of Cooper Pairs in a Mesoscopic 2D Fermi Gas}
\author{Marvin~Holten}
    \email{mholten@physi.uni-heidelberg.de}
	\affiliation{Physikalisches Institut der Universit\"at Heidelberg, Im Neuenheimer Feld 226, 69120 Heidelberg, Germany}
\author{Luca~Bayha}
	\affiliation{Physikalisches Institut der Universit\"at Heidelberg, Im Neuenheimer Feld 226, 69120 Heidelberg, Germany}
\author{Keerthan~Subramanian}
	\affiliation{Physikalisches Institut der Universit\"at Heidelberg, Im Neuenheimer Feld 226, 69120 Heidelberg, Germany}
\author{Sandra~Brandstetter}
	\affiliation{Physikalisches Institut der Universit\"at Heidelberg, Im Neuenheimer Feld 226, 69120 Heidelberg, Germany}
\author{Carl~Heintze}
	\affiliation{Physikalisches Institut der Universit\"at Heidelberg, Im Neuenheimer Feld 226, 69120 Heidelberg, Germany}
\author{Philipp~Lunt}
	\affiliation{Physikalisches Institut der Universit\"at Heidelberg, Im Neuenheimer Feld 226, 69120 Heidelberg, Germany}
\author{Philipp~M.~Preiss}
	\affiliation{Physikalisches Institut der Universit\"at Heidelberg, Im Neuenheimer Feld 226, 69120 Heidelberg, Germany}
	\affiliation{Present Address: Max Planck Institute of Quantum Optics, Hans-Kopfermann-Str.\ 1, 85748 Garching, Germany}
\author{Selim~Jochim}
	\affiliation{Physikalisches Institut der Universit\"at Heidelberg, Im Neuenheimer Feld 226, 69120 Heidelberg, Germany}

\date{\today}

\blfootnote{\href{mailto:mholten@physi.uni-heidelberg.de}{$^\ast$ mholten@physi.uni-heidelberg.de}}
\maketitle
%%%%%%%%%%%%%%%%%%%%%%%%%%%%%%%%%%%%%%%%%%%%%%%%%%%%%%%%%%%%%%%%%%%%%%%%%%%%%%%%%
%						        	Abstract        	   						%
%%%%%%%%%%%%%%%%%%%%%%%%%%%%%%%%%%%%%%%%%%%%%%%%%%%%%%%%%%%%%%%%%%%%%%%%%%%%%%%%%
\textbf{
Pairing is the fundamental requirement for fermionic superfluidity and superconductivity \cite{Yang_1962}. 
To understand the mechanism behind pair formation is an ongoing challenge in the study of many strongly correlated fermionic systems \cite{Zhou_2021}.
Cooper pairs are the key ingredient to BCS theory as the microscopic explanation of conventional superconductivity \cite{Bardeen_1957}.
They form between particles of opposite spin and momentum at the Fermi surface of the system.
Here, we directly observe Cooper pairs in a mesoscopic two-dimensional Fermi gas.
We apply an imaging scheme that enables us to extract the full in-situ momentum distribution of a strongly interacting Fermi gas with single particle and spin resolution \cite{Holten_2021}.
Our ultracold gas allows us to freely tune between a completely non-interacting, unpaired system and weak attractions, where we find Cooper pair correlations at the Fermi surface.
When increasing the attractive interactions even further, the pairs gradually turn into deeply bound molecules breaking up the Fermi surface.
Our mesoscopic system is closely related to the physics of nuclei, superconducting grains or quantum dots \cite{Alhassid_2000,Delft_2001,Launey_2017}.
With the precise control over interactions, particle number and potential landscape in our experiment, the observables we establish in this work provide a new approach to longstanding questions concerning not only such mesoscopic systems but also their connection to the macroscopic world.
}

%%%%%%%%%%%%%%%%%%%%%%%%%%%%%%%%%%%%%%%%%%%%%%%%%%%%%%%%%%%%%%%%%%%%%%%%%%%%%%%%%
%						        	Main Text         	   						%
%%%%%%%%%%%%%%%%%%%%%%%%%%%%%%%%%%%%%%%%%%%%%%%%%%%%%%%%%%%%%%%%%%%%%%%%%%%%%%%%%
Quantum states can be characterized by detecting all the correlations between the constituents of the system \cite{Altman_2004,Schweigler_2017}.
However, the amount of information that is available scales exponentially with the size of the system \cite{Flammia_2012}.
A crucial challenge in the study of quantum many-body systems is to identify and detect the relevant correlations that efficiently describe a state of matter \cite{Zache_2020}.
For example, it took more than forty years after the discovery of conventional superconductors before Bardeen, Cooper and Schrieffer came up with their explanation in terms of bound electrons or Cooper pairs \cite{Bardeen_1957}.
These are correlations between pairs of fermionic particles with opposite momentum that are localized at the Fermi surface in momentum space \cite{Cooper_1956}.
And while it was quickly understood that pairing is the key ingredient for fermionic superfluidity and superconductivity \cite{Yang_1962}, for many systems, most prominently high-$T_\text{c}$ superconductors, the exact nature of correlations remains unknown \cite{Lee_2006,Zhou_2021}.

Ultracold quantum gases are an ideal platform for the simulation and study of strongly correlated Fermi superfluids in this context.
They offer a unique setting with full tunability of interactions, particle numbers and single particle spectra combined with high-fidelity detection methods \cite{Bloch_2008,Bloch_2012}.
Density and spin correlations, for example, can be accessed directly in the atomic noise in an image of an expanding gas, even without single atom resolution \cite{Altman_2004}.
This method has been applied successfully to both bosonic and fermionic quantum gases \cite{Greiner_2005,Foelling_2005,Rom_2006,Spielman_2007,Jeltes_2007,Tenart_2021}.
For lattice systems, quantum gas microscopy has become an increasingly powerful tool to study spatial correlations at the microscopic level \cite{Bakr_2009,Sherson_2010,Parsons_2016,Koepsell_2020}.

In this work, we study the emergence of fermionic pair correlations in momentum space in spatially continuous two dimensional (2D) systems starting from the smallest possible instance.
Our fluorescence imaging technique allows us to extract the spin and single atom resolved momentum distribution with particle detection fidelities comparable to those typically achieved in quantum gas microscopes \cite{Bergschneider_2018, Preiss_2019} (see Methods).
Previously, we have established this method for small systems of indistinguishable Fermions and found strong correlations in their relative positions as a manifestation of Pauli's principle \cite{Holten_2021}.
Here, we study a two-component Fermi gas with 12 particles trapped in a 2D harmonic potential and with freely tunable attractive interactions.
The particles are prepared in the closed-shell ground state configurations of the harmonic oscillator with high-fidelity \cite{Bayha_2020}.

Our measurements enable us to extract the pair correlations and paired fraction as a function of attraction strength.
This allows us to directly identify Cooper pairs emerging at the Fermi surface as the relevant correlations for small attraction strengths.
At much stronger interactions we find pair correlations also inside the Fermi sea, indicating a transition to molecular pairing.
Our work pioneers the single particle resolved study of correlations in momentum space.
It lays the foundation for future studies also in more complex settings with more particles, imbalance or using higher temperature states, for example in the strongly correlated region of the BEC-BCS crossover.

\begin{figure}
	\centering
	\includegraphics{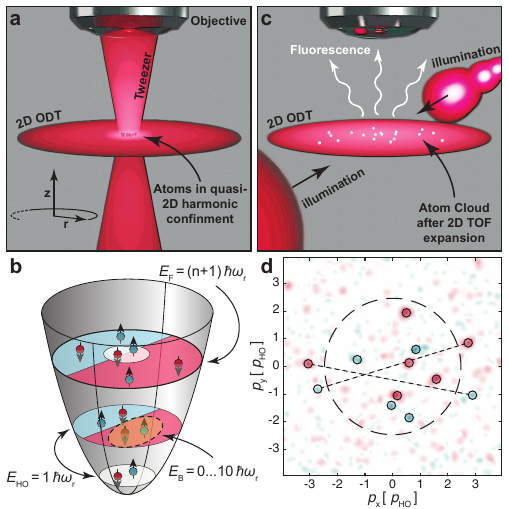}
	\caption{\textbf{Preparation and detection of a mesoscopic 2D Fermi gas.} \textbf{a,} Sketch of the experimental setup. The fermionic atoms are trapped in the center of a potential created by superimposing an optical tweezer with a 2D optical dipole trap (2D ODT). \textbf{b,} The degeneracy of the 2D harmonic oscillator potential leads to stable configurations when filled with 1+1, 3+3, 6+6 or 10+10 atoms. The behaviour of these closed-shell systems is determined by the ratio between the three intrinsic energy scales of the system. These are the single particle gap $E_\text{HO}=\SI{1}{\hbar\omega_\text{r}}$, the Fermi energy $E_\text{F}=(n+1)\,\hbar\omega_\text{r}$ and the binding energy $E_\text{B}$ that we control with an magnetic offset field. \textbf{c,} Sketch of the imaging scheme. To measure the momentum distribution we let the cloud expand in the 2D dipole trap. The atoms are imaged through the microscope objective by exciting them with two resonant counterpropagating illumination beams. \textbf{d,} Single image of the momentum distribution of the 6+6 atom ground state. The dashed black circle indicates the Fermi momentum $p_\text{F}$. The highlighted particles contribute to the correlation peak that we observe at the Fermi surface at any attraction strengths greater than zero. The finite size of our sample leads to fluctuations of the center of mass momentum of the pairs around zero. As a result, the lines connecting the pairs do not go exactly through the center of the circle.}
	\label{fig:mainSetup}
\end{figure}

\section{Experimental Setup}
We start our experiments with a balanced mixture of two hyperfine states of $^6$Li atoms.
An optical tweezer (OT) is providing an approximately harmonic confinement in radial direction with frequency $\omega_\text{r} = 2 \pi \times \SI{1101 \pm 2}{\Hz}$.
To achieve a quasi-2D confinement, we superimpose the OT with a single anti-node of an optical lattice (2D ODT) in vertical direction, providing an axial confinement of $\omega_z = 2\pi \times \SI{7432 \pm 3}{\Hz}$.
This results in an oblate trapping geometry with an aspect ratio of $7:1$, where the atoms remain in the axial ground state (see Fig.\,\ref{fig:mainSetup}a).
A spilling technique allows us to deterministically prepare the ground state of up to $N=10$ atoms per spin component (denoted as 10+10) in this potential and with high-fidelities \cite{Serwane_2011}.
For 6+6 atoms, for example, we prepare the ground state in $\SI{76 \pm 2}{\%}$ of the experimental cycles and estimate that the remaining entropy per particle is approximately $\SI{0.1}{k_\text{B}}$ \cite{Bayha_2020}.
The degeneracy of the $n$th level of the 2D harmonic oscillator is $n+1$ ($n=0,1,...$).
This leads to stable closed-shell configurations for the ground states of 1+1, 3+3, 6+6 and 10+10 atoms (see Fig.\,\ref{fig:mainSetup}b) and with Fermi energies of $E_\text{F}=(n_\text{F}+1)\hbar\omega_\text{r}$.
Here, $n_\text{F}=(\sqrt{2N+1/4}-3/2)$ denotes the highest completely filled oscillator level and $N$ is the single-spin atom number.
The tightly focused OT results in an anharmonicity of the 2D potential and reduced trap frequencies ($\sim \SI{10}{\%}$) for larger shell numbers ($n\gtrsim 2$).

We control the interactions between the particles by applying a magnetic offset field and using a Feshbach resonance \cite{Zuern_2013}.
In a 2D geometry the interaction properties are fully determined by a bound state of energy $E_\text{B}$ that is present for any attractive contact interaction strength \cite{Randeria_1990}.
We express $E_\text{B}$ in units of the harmonic oscillator frequency in radial direction $\hbar\omega_\text{r}$.
We ensure that the binding energy is always smaller than the axial confinement $E_\text{B} < \hbar \omega_\text{z}$ to remain in the quasi-2D limit.

\begin{figure*}
	\centering
	\includegraphics{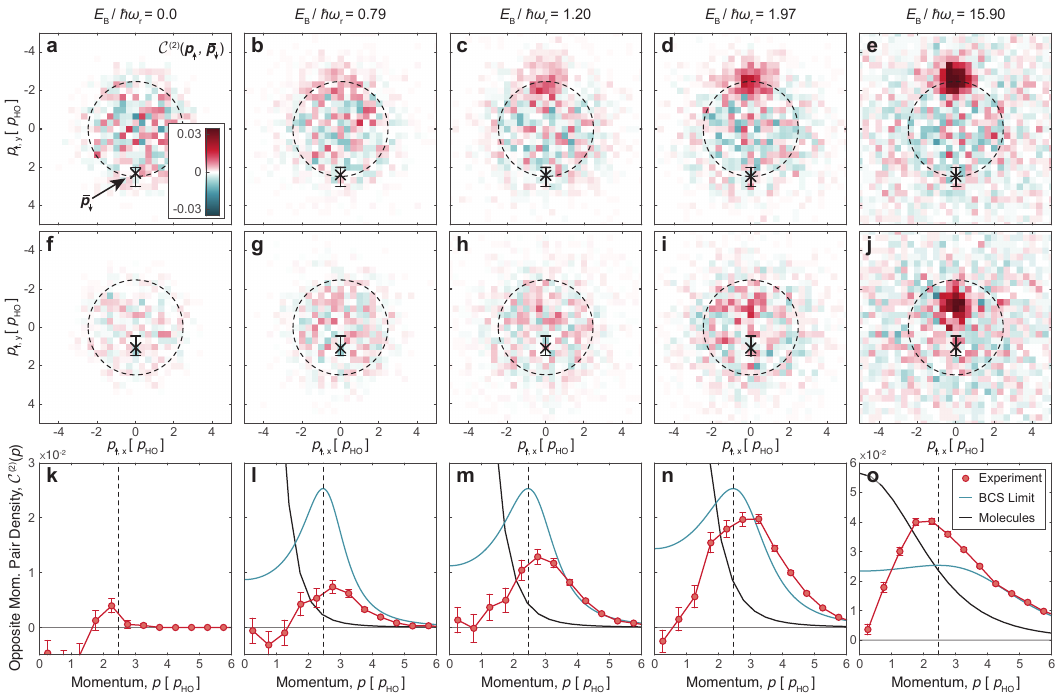}
	\caption{
		\textbf{Density-density correlation function.}
		\textbf{a-j,} The density plots show $\mathcal{C}_{\overline{p}_\downarrow}^{(2)}$, the normalized probability of detecting a spin up particle at momentum $\bm{p}_\uparrow$, given that a spin down particle was detected with the momentum $\bm{\overline{p}}_\downarrow$.
		The cross shows the mean momentum of the spin down atom $\overline{p}_\downarrow$, averaged over all images that contribute to the selected momentum bin (i.e. where a spin down particle is found between the horizontal bars).
		The dashed circle marks the Fermi momentum, defined as $p_\text{F}=\sqrt{6}\, p_\text{HO}$.
		\textbf{a-e,} The reference particle is located inside the Fermi sea {($\overline{p}_\downarrow$ between $p_1=0.5$ and $p_2=1.5\, p_\text{HO}$)}.
		Significant correlations appear only at large binding energies ($E_\text{B}/E_\text{F}\gg 1 $)
		\textbf{f-j,} Positive correlations are visible already at $E_\text{B}/\hbar\omega_\text{r} \approx 1$, when the reference particle is placed at the Fermi surface instead ($p_1 = 2$ to $p_2=3\,p_\text{HO}$).
		They are located opposite to the spin down particle in momentum space $\Delta\phi = \pi$ and, in the weak binding regime, present only at the Fermi surface (b,g and c,h).
		This identifies them as Cooper pairs.
		\textbf{k-o,} The opposite momentum pair density $\mathcal{C}^{(2)}(\bm{p},-\bm{p})$ is shown.
		BCS theory (solid blue line) correctly predicts the appearance of a correlation peak at $p_\text{F}$.
		For a system of non-interacting 2D molecules in the ground state, we expect a very strong correlation peak centered around zero momentum (solid black line).
		The error bars represent the standard error of the mean.
	}
	\label{fig:mainC2}
\end{figure*}

\section{Single Particle Imaging}
High resolution fluorescence imaging allows us to extract the in-situ momentum distribution of any quantum state we can prepare and at any interaction strength.
First, we map the in-situ momentum of each particle onto its position by a non-interacting time of flight (TOF) expansion in 2D (see Fig.\,\ref{fig:mainSetup}c).
Subsequently, the position of each particle is recorded on a camera with a high-fidelity fluorescence imaging scheme \cite{Bergschneider_2018} (see Fig.\,\ref{fig:mainSetup}d).
We achieve single atom detection fidelities in the complete field of view of $\SI{97.8\pm 0.9}{\%}$ (see Methods).
To ensure that scattering events during the expansion do not alter the measured momentum distribution, even when starting from a strongly interacting state, we switch off all interactions at the beginning of the TOF sequence.
To this end, we transfer all the atoms in one of the two spin components into a third hyperfine state that does not interact with any of the initial state atoms.
The projection is driven by two copropagating Raman laser beams and is almost three orders of magnitude faster ($T_{\pi}=\SI{300}{\ns}$) than any other intrinsic timescale of our many-body state and can therefore be considered as instantaneous projection (see Methods).
All the momenta are expressed in natural units of the in-situ harmonic oscillator potential given by $p_\text{HO} = \sqrt{\hbar m \omega_\text{r}}$, where $m$ is the mass of the $^6$Li atoms.
We define the Fermi momentum as $p_\text{F}=\sqrt{2mE_\text{F}}=\sqrt{2(n_\text{F}+1)}p_\text{HO}$ (see Methods).

\section{Momentum Correlations}
Each image represents a single projection of the full many-body wavefunction.
We take $\SI{1000}{}$ of these snapshots of the 6+6 atom ground state (see Fig.\,\ref{fig:mainSetup}d) for each interaction strength and search for the relevant correlations.
To study fermionic pairing, a natural choice is the opposite spin, density-density correlator $\mathcal{C}^{(2)}$, defined as:
\begin{equation}\label{eqn:c2}
    \mathcal{C}^{(2)}(\bm{p}_\uparrow,\bm{p}_\downarrow) = \langle n(\bm{p}_\uparrow)n(\bm{p}_\downarrow) \rangle - \langle n(\bm{p}_\uparrow) \rangle \langle n(\bm{p}_\downarrow) \rangle.
\end{equation}
Here, $n$ denotes the density operator and $\langle ... \rangle$ is the average over all images.
The correlation function $\mathcal{C}^{(2)}$ expresses the conditional probability of finding a spin up particle at momentum $\bm{p}_\uparrow$ given that a reference particle with spin down was detected at $\bm{p}_\downarrow$ after subtracting the contribution from single particle densities $\langle n(\bm{p}_\uparrow) \rangle \langle n(\bm{p}_\downarrow) \rangle$.

Since $\mathcal{C}^{(2)}$ depends on four coordinates, additional steps are required to display the relevant correlations.
We fix the reference spin down particle to some momentum $\bm{\overline{p}}_\downarrow$ and plot the $C(2)$ as a function of $\bm{p}_\uparrow$.
A binning procedure is required to extract the correlation function from the discrete experimental measurements.
It is convenient to perform the binning in polar coordinates $\bm{p_i} \rightarrow (p_i,\phi_i)$.
In Fig.\,\ref{fig:mainC2} a-j we show 
\begin{equation}\label{eqn:c2_int}
	\begin{aligned}
   &\mathcal{C}_{\overline{p}_\downarrow}^{(2)}(p_\uparrow,\Delta \phi )= \int_{p_1}^{p_2}\int_{0}^{2\pi}\int_{0}^{2\pi} \diff p'_\downarrow \diff \phi'_\downarrow \diff \phi'_\uparrow \\ &\quad \mathcal{C}^{(2)}(p_\uparrow,\phi'_\uparrow,p'_\downarrow,\phi'_\downarrow)\,\delta(\Delta \phi - (\phi'_\uparrow-\phi'_\downarrow)) \, p'_\downarrow,
	\end{aligned}
\end{equation}
where we integrate the momentum of the reference spin down particle $\overline{p}_\downarrow$ over a bin size of $\Delta p = p_2-p_1 = 1\,p_\text{HO}$ (indicated by the horizontal bars).
The black cross indicates the mean momentum of the spin down atom $\overline{p}_\downarrow$, averaged over all measurements that contribute to the integral.
The integrals over the angle take advantage of the radial symmetry of the system and average over all points in the correlator with the same relative angle $\Delta \phi = \phi_\uparrow-\phi_\downarrow$ (for more details see Methods).
The density plots of $\mathcal{C}_{\overline{p}_\downarrow}^{(2)}$ visualize at what momentum $\bm{p}_\uparrow$ the probability of detecting spin up particles is enhanced when a spin down particle is present at the momentum $\overline{p}_\downarrow$.

The measurements reveal how pairing emerges in the ground state of the mesoscopic gas as the attraction strength is increased.
Inside the Fermi sea, correlations are strongly suppressed when $E_\text{B} \lesssim E_\text{F}$ (see Fig.\,\ref{fig:mainC2}a-d).
At the Fermi surface however, a clear correlation peak appears as soon as the binding energy is on the order of the single particle gap $E_\text{B} \gtrsim E_\text{HO} = \hbar\omega_\text{r}$ (f-i).
The correlation peak increases with binding energy and is strongest at the Fermi momentum $p_\text{F}$ and at $\Delta \phi = \pi$ directly opposite to the reference spin down particle.
This demonstrates that we can observe Cooper pairs directly in a strongly correlated mesoscopic Fermi gas in 2D.
By increasing the binding energy much further to $E_\text{B} \gg E_\text{F}$ we are able to enter the regime of more tightly bound dimers where pair correlations emerge also inside the Fermi sea (e,j).
For the data points in the molecular regime, we have reduced $\omega_\text{r}$ to $2 \pi \times \SI{343 \pm 5}{\Hz}$ to ensure that the condition $E_\text{B} < \hbar \omega_\text{z}$ remains fulfilled.

\section{Emergence of Pairing}
\begin{figure}
	\centering
	\includegraphics{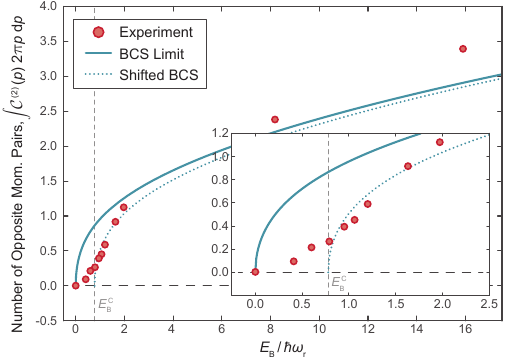}
	\caption{
		\textbf{Number of opposite momentum pairs as a function of the interaction strength.}
		We define the number of opposite momentum pairs at the 2D integral of $\mathcal{C}^{(2)}(p,-p)$ over the momentum $p$.
		As a result of the finite gap $\hbar\omega_\text{r}$, the transition from a unpaired to a paired ground state is expected to occur at a critical value for the binding strength $E_\text{B}^\text{C}=0.78\hbar\omega_\text{r}$, even at zero temperature.
		As a first approximation of this finite size effect, we shift the prediction from standard BCS theory (solid line) by $E_\text{B}^\text{C}$ (dotted line).
		The inset shows the weakly paired region in more detail.
		The small and fixed particle number, as opposed to a grand canonical ensemble, explain the smooth transition.
		The statistical errors are on the order of the symbol size.}
	\label{fig:mainPairs}
\end{figure}

Between different spins there are significant second-order correlations only between particles with opposite momenta in our system (see Fig.\,\ref{fig:mainC2}a-j).
To get a more quantitative picture of how pairing emerges, we therefore extract the opposite momentum pair density $\mathcal{C}^{(2)}(\bm{p}_\uparrow\rightarrow \bm{p},\bm{p}_\downarrow\rightarrow -\bm{p})$, as defined by equation \ref{eqn:c2} (see Fig.\,\ref{fig:mainC2}k-o).
Due to the radial symmetry of the system, $\mathcal{C}^{(2)}(\bm{p},-\bm{p})\equiv \mathcal{C}^{(2)}(p)$ depends only on the magnitude of $\bm{p}$.
The total number of pairs in the ground state can then be extracted by integrating over $\mathcal{C}^{(2)}(p)$ (see Fig.\,\ref{fig:mainPairs}).
The measurements reveal how the ground state transforms from the non-interacting, completely unpaired state to a paired system.

We identify three different regimes of pair correlations.
The weakly paired regime $E_\text{B}\lesssim \hbar\omega_\text{r}$, the regime of intermediate interaction strength $E_\text{B} \sim E_\text{F}\gtrsim \hbar\omega_\text{r}$ and the limit of strong binding $E_\text{B}\gg E_\text{F}$.
In the regime of weak interactions only a small fraction of the system shows pair correlations.
For the largest accessible binding energies of $E_\text{B}= 15.9\hbar\omega_{r}$ the number of total pairs is $\SI{4.1\pm0.1}{}$, closer to the maximum possible value of 6 for the 6+6 particle system (see Fig.\,\ref{fig:mainPairs}).
Here, the interactions between the bosonic pairs are still very large and this measurement is still in the strongly correlated regime of intermediate interactions.
To reach the strong binding limit described by point-like molecules without interactions we would have to increase $E_\text{B}$ even further.

We compare the measurements to standard BCS theory and to a model system of N=6 non-interacting 2D dimers in the harmonic oscillator ground state.
We expect these mean-field descriptions to become accurate when $N\rightarrow \infty$ and in the limits of weak and strong binding respectively.
BCS theory can qualitatively explain the presence of a correlation peak at the Fermi surface that we find as the main feature of our system in the experiment (see Fig.\,\ref{fig:mainC2}l-n).
For binding energies much larger than the single particle gap $E_\text{B}\gg \hbar\omega_\text{r}$, we find that the correlations become much stronger and their maximum shifts towards smaller momenta (see Fig.\,\ref{fig:mainC2}o).
This qualitatively agrees with the expectation for a system of tightly bound 2D dimers in the ground state of the harmonic oscillator where the correlation peak is centered around zero momentum (solid black line).
Both mean-field descriptions generally fail to produce accurate quantitative predictions indicating that both beyond mean-field and finite size effects are present in our experiment.

The mesoscopic Fermi gas in a 2D harmonic oscillator is closely related to superconducting grains, quantum dots and systems from nuclear and atomic physics \cite{Bohr_1975,Alhassid_2000,Delft_2001,Launey_2017}.
When the coherence length approaches the system size, quantum confinement effects become important and lead to a discrete single particle spectrum.
As soon as the level spacing becomes of the order of the many-body gap $\Delta$ superfluity breaks down and the system remains in the normal phase, even at zero temperature \cite{Delft_2001}.
Due to the small and fixed particle number and discrete spectrum, mean-field approaches generally break down and descriptions in terms of local quantities, like the conductance, become impossible.
Instead the sample has to be treated as a whole.
In the closed-shell configurations we study in our experiment, all the levels up to the Fermi energy $E_\text{F}$ are already occupied and there is a gap of $1\,\hbar\omega_\text{r}$ to the next unoccupied levels.
In the thermodynamic limit, when $N\rightarrow\infty$, this leads to a phase transition from a normal to a superfluid phase at some critical value for the binding energy $E_\text{B}^\text{C}$ \cite{Bruun_2014}.
A precursor of this phase transition can be observed already at the mesoscopic scale \cite{Bjerlin_2016,Bayha_2020}.
The critical value for N=6+6 particles is predicted as $E_\text{B}^\text{C}=0.78\hbar\omega_\text{r}$ from an exact diagonalization of the Hamiltonian \cite{Bjerlin_2016}.

In Fig.\,3 we plot BCS theory shifted by the critical binding energy $E_\text{B}^{C}$ as a first order approximation of the finite size effects (dotted line).
In the weakly paired regime this explains the increase of the pair number of the closed-shell ground state as a function of the binding energy $E_\text{B}$ (inset).
Due to the small particle number the transition is much smoother than the sharp increase at $E_\text{B}^{C}$ that is expected for larger systems.
The large single particle gap allows us to study the weakly paired regime at much larger absolute binding energies $E_\text{B}\sim \hbar\omega_\text{r}$ and temperatures than what would be required for macroscopic samples.
When the attraction strength is increased further we enter the strongly correlated regime and the measured number of pairs increases above the mean-field prediction (see Fig.\,3).
Here, fluctuations of the many-body gap beyond the mean-field value $\Delta$ have to be considered for a more accurate quantitative prediction \cite{Randeria_1989}

When the particle number is increased $N\rightarrow\infty$, we expect the correlations to become even sharper peaked around the Fermi surface in the weakly interacting limit.
The limiting cases of infinite and weak attraction converge against the mean-field description.
In the regime of intermediate interactions, new theories are required for a quantitatively accurate prediction of the pair correlations.
The precise measurements of correlations, also beyond second order, in our experiment can be used as important benchmarks for new numerical and analytical approaches in the future.

\section{Outlook}
In conclusion, we have demonstrated that single particle resolved correlations can be accessed in continuous systems that are strongly interacting.
We directly observe how Cooper pairs emerge at the Fermi surface.
The correlation signal as a function of momentum and attraction strength allows us to characterize the ground state and identify different pairing mechanisms.
An even more thorough characterization of quantum many body states will become possible when extending our imaging scheme to detect the single particle resolved in-situ wavefunction in addition \cite{Asteria_2021}.
Our experiment opens up new pathways for detailed studies of the normal phase of the BEC-BCS crossover \cite{Murthy_2017}, imbalanced systems in 1D or 2D \cite{Pecak_2020,Chevy_2010} or rotating systems \cite{Palm_2020,Pecci_2021}.
By unlocking the capability to access correlations of arbitrary order, our method hold the potential to gain deeper insight into such strongly correlated systems in an unprecedented way.

%%%%%%%%%%%%%%%%%%%%%%%%%%%%%%%%%%%%%%%%%%%%%%%%%%%%%%%%%%%%%%%%%%%%%%%%%%%%%%%%%
%					     Acknowledgements and Contributions   		   			%
%%%%%%%%%%%%%%%%%%%%%%%%%%%%%%%%%%%%%%%%%%%%%%%%%%%%%%%%%%%%%%%%%%%%%%%%%%%%%%%%%

\paragraph*{Data availability}
The data that support the findings of this study are available from the corresponding authors upon request. Source data for Figures 2k-o and 3 are provided with this paper.

\paragraph*{Acknowledgements}
We gratefully acknowledge insightful discussions with A.\ Faribault, J.\ von Delft.
This work has been supported in parts by the Deutsche Forschungsgemeinschaft (DFG, German Research Foundation): the Collaborative Research Centre SFB 1225 (ISOQUANT) and Germany’s Excellence Strategy EXC2181/1-390900948 (the Heidelberg STRUCTURES Excellence Cluster).
It was also funded by the European Union’s Horizon 2020 research and innovation program under grant agreements No.~817482 PASQuanS and grant No.~725636 (QuStA).
Support by the Heidelberg Center for Quantum Dynamics is acknowledged.

\paragraph*{Author Contributions}
M.H.,\ and L.B.\ and K.S.\ performed the measurements. M.H.,\ L.B.,\ K.S.\ and S.B.\ analyzed the data. C.H.\ and M.H.\ performed the numerical calculations. P.L. set up the phase-locked loop for the Raman beams. P.M.P.\ and S.J.\ supervised the project. M.H. wrote the manuscript with input from all authors. All authors contributed to the discussion of the results.

\paragraph*{Competing Interest}
The authors declare no competing interests.

\paragraph*{Correspondence and requests for materials}
should be addressed to M.H.

%%%%%%%%%%%%%%%%%%%%%%%%%%%%%%%%%%%%%%%%%%%%%%%%%%%%%%%%%%%%%%%%%%%%%%%%%%%%%%%%%
%					     Bibliography                          		   			%
%%%%%%%%%%%%%%%%%%%%%%%%%%%%%%%%%%%%%%%%%%%%%%%%%%%%%%%%%%%%%%%%%%%%%%%%%%%%%%%%%

%%%%%%%%%%%%%%%%%%%%%%%%%%%%%%%%%%%%%%%%%%%%%%%%%%%%%%%%%%%%%%%%%%%%%%%%%%%%%%%%%
%								Supplemental Material							%
%%%%%%%%%%%%%%%%%%%%%%%%%%%%%%%%%%%%%%%%%%%%%%%%%%%%%%%%%%%%%%%%%%%%%%%%%%%%%%%%%

\setcounter{figure}{0}
\renewcommand{\figurename}{Extended Data Figure}
\cleardoublepage
\newpage
\section*{Methods}
\begin{figure*}
    \centering
	\includegraphics{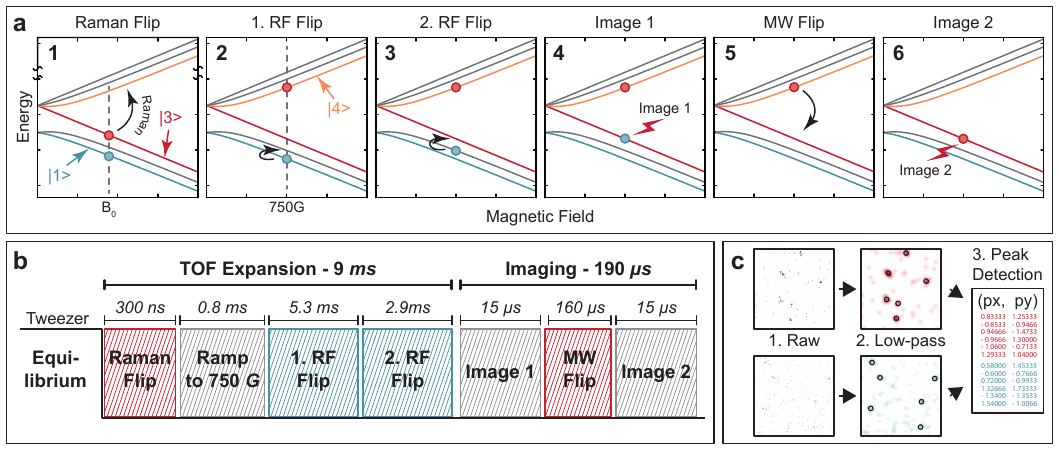}
    \caption{ \textbf{Sketch of the TOF imaging sequence.} \textbf{a,} The TOF imaging scheme is initiated by switching off the trapping potential and the interparticle interactions. To this end, we shine in two copropagating Raman laser beams to quasi-instantaneously transfer all atoms from state $\ket{3}$ to state $\ket{4}$ (1). After the Raman transition, we ramp the magnetic field from the value $B_\text{0}$ that sets the in-situ interaction strength to a constant value of $B=\SI{750}{\gauss}$ where the fidelity of the following spin flips is maximal. Only the imaging transitions of state $\ket{3}$ and $\ket{6}$ are closed. Two successive radio frequency (RF) Landau-Zener sweeps are applied to move all the atoms from state $\ket{1}$ to $\ket{3}$ during the free expansion (2, 3). This is followed by taking the first image with illumination beams that are resonant only to state $\ket{3}$ (4). Before taking the second image, a microwave (MW) Landau-Zener sweep transfers the remaining atoms in $\ket{4}$ to state $\ket{3}$ again (5,6). The MW pulse leads to higher transfer fidelities than the Raman lasers but is on the other hand much slower. \textbf{b,} The duration of the initial Raman flip with $T_\pi=\SI{300}{\ns}$ is chosen as fast as technically possible to prevent any scattering between atoms from occurring during the expansion. The RF Flips are solely optimized for transfer fidelities and are distributed over the remaining time of $T_\text{TOF}=\SI{9}{\ms}$. The time of the MW flip is set by the maximal frame rate of the camera. \textbf{c,} From a single experimental run we obtain two binary images where bright pixel indicate that at least one photon hit the chip at this location. In the first step we apply a low-pass filter to these images. A simple peak detection with an optimized acceptance threshold then allows us to extract the position of all spin up and down particles respectively.}
    \label{fig:imSeq}
\end{figure*}

\paragraph*{\textbf{Preparation Sequence}}
A more detailed explanation of the experimental sequence can be found in ref. \cite{Bayha_2020}.
We apply the same scheme here to prepare the closed-shell ground state configurations of up to 10+10 particles of the 2D harmonic oscillator. 
We start by transferring a cold gas of \textsuperscript{6}Li atoms from a magneto-optical trap into a red-detuned optical dipole trap (ODT).
A radio frequency (RF) pulse sequence is applied to create a balanced mixture of atoms in hyperfine states $\ket{1}$, $\ket{3}$.
Here, we label the hyperfine states of the \textsuperscript{2}S\textsubscript{1/2} ground state of \textsuperscript{6}Li according to their energy in increasing order from $\ket{1}$ to $\ket{6}$ (see Extended Data Fig.\,\ref{fig:imSeq}a).
Next, a tightly focused optical tweezer (OT) is loaded from the ODT and several evaporation stages are used to create a deeply degenerate gas of around 350 atoms in the OT.
A precise spilling method, discussed in detail in \cite{Serwane_2011}, results in about 30 atoms in the OT with all levels up to the Fermi surface filled with very high probabilities.

\begin{figure}
    \centering
	\includegraphics{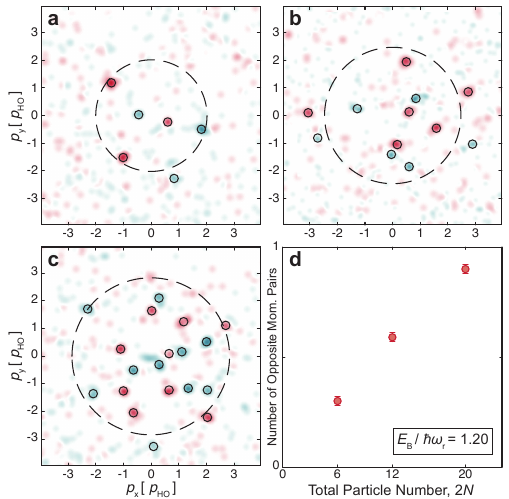}
    \caption{ \textbf{Different particle numbers.} In our experiment, we are able to prepare closed- and open-shell configurations with different particle numbers. \textbf{a-c,} Single momentum space projections of the three lowest closed-shell ground states with 3+3, 6+6 and 10+10 particles respectively are shown. The dashed circle indicates the corresponding Fermi momentum for each particle number. \textbf{d,} The total weight in the pair correlation peak of $\mathcal{C}_\uparrow^{(2)}$ is plotted versus particle number and for the setting where the reference particle is fixed at the Fermi surface (for 6+6 see Fig.\,\ref{fig:mainC2}h). We find that the absolute number of paired atoms increases while their fractions remains approximately constant from N=3+3 to N=10+10 particles.
    The error bars represent the standard error of the mean.}
    \label{fig:N}
\end{figure}

The OT has a quasi-1D aspect ratio of $\omega_r:\omega_z = 5:1$.
To create a 2D sample, we perform a continuous crossover to a quasi-2D confinement with an aspect ratio of $\omega_r:\omega_z = 1:7$.
To this end, we ramp up the power of a 2D dipole trap with a trap frequency of $\omega_z = 2\pi \times \SI{7432\pm 3}{\hertz}$ and aspect ratio of $\omega_r:\omega_z\approx1:300$.
At the same time, the radial frequency of the OT is reduced from $\omega_r \approx 2\pi \times \SI{20}{\kilo\hertz}$ to $2\pi \times \SI{1101\pm2}{\hertz}$ by increasing the beam waist from about $\SI{1}{\micro \meter}$ to $\SI{5}{\micro \meter}$ with a spatial light modulator.
The final potential configuration is shown in Fig.\,\ref{fig:mainSetup}a.
In the combined quasi-2D potential we are able to reach the closed-shell ground state configurations of the four lowest shells of the 2D harmonic oscillator with a final high-fidelity spilling sequence (see Extended Data Fig. \ref{fig:N}).
After the ground state has been initialized, the interaction strength is set by adiabatically ramping the magnetic offset field $B_\text{0}$ to obtain the desired value for $E_\text{B}/ \hbar \omega_\text{r}$.

\paragraph*{\textbf{Imaging Sequence}}
A detailed sketch of the imaging sequence is shown in Extended Data Fig.\,\ref{fig:imSeq}.
It can be separated into two parts.
First, the free time of flight (TOF) expansion where the cloud size increases by a factor of approximately 50 and the in-situ momenta of the particles are mapped onto their position.
Second, the image acquisition itself, where the atoms are excited by resonant laser beams, start to fluoresce and their positions are recorded on a camera.

The TOF sequence begins after the ground state with the desired particle number and interaction strength has been prepared.
The first step is to switch off the OT which radially confines the cloud.
This leads to a quick expansion of the gas in the 2D dipole trap (see Fig.\,\ref{fig:mainSetup}a,c).
The expansion serves two purposes.
The increased distance between the single atoms allows us to resolve them in the first place.
Second, it maps the in-situ momentum of each particle onto its position.
For the study of fermionic superfluidity, imaging in momentum space is advantageous over position space since it offers more direct access to the relevant correlations --- like Cooper pairs --- of the gas.
In the future, we plan to extend our scheme to enable us to take images of the in-situ density with single particle resolution in addition \cite{Asteria_2021}.

To ensure that the mapping into momentum space by the TOF is accurate and that we measure genuine in-situ correlations, no scattering events may occur during the expansion.
To this end, we use two Raman laser beams to quickly switch from the strongly interacting $\ket{1} - \ket{3}$ to the almost non-interacting $\ket{1} - \ket{4}$ mixture in $T_\pi=\SI{300}{\ns}$ (see Extended Data Fig.\,\ref{fig:imSeq}a,b).
Between states $\ket{1}$ and $\ket{4}$ there is no Feshbach resonance present and all our measurements are consistent with their scattering length $a_\text{14}$ being very close to zero.
By checking for interaction shifts in the spectrum of two particles in a harmonic trap we determined an upper limit for the scattering length as $\left| a_\text{14} \right|< 500 a_\text{0}$.
This sets an upper limit of one scattering event between two atoms in 50 experimental runs for our parameters when expanding in the $\ket{1}-\ket{4}$ mixture.

\begin{figure}
    \centering
	\includegraphics{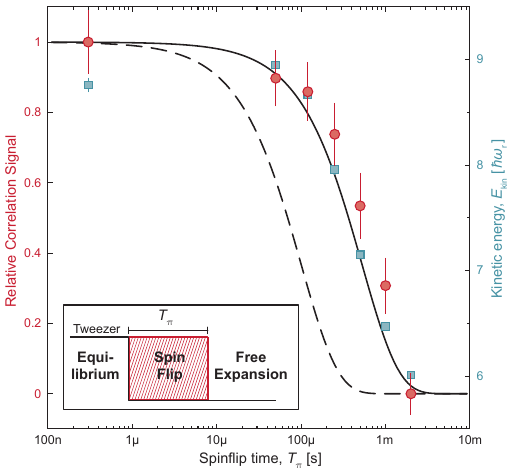}
    \caption{ \textbf{Scanning the interaction switchoff time.} $T_\pi$ is the duration of the spin flip from a strongly interacting $\ket{1}-\ket{3}$ to an (almost) non-interacting $\ket{1}-\ket{4}$ mixture at the beginning of the TOF sequence (inset). When we increase $T_\pi$, the magnitude of the pair correlations reduce significantly above a threshold of $T_\pi= \SI{100}{\us}$ (red circles). The reason is that scattering events during the TOF expansion redistribute the momenta between the participating atoms and destroy the correlations that were present between the in-situ momenta. We model the effect by assuming that each scattering event between two-atoms annihilates all the in-situ correlations for those particles. The only parameters that enter the model (solid line) are the scattering rate $\lambda_\text{sc}$ of the $\ket{1}-\ket{3}$ mixture at the magnetic offset field ($B_\text{0}=\SI{750}{\gauss}$) and the in-situ density. The dashed line shows the same model prediction but at one of the highest scattering rates used in our experiments ($B_\text{0}=\SI{695}{\gauss}$, $E_\text{B}/\hbar\omega_\text{r}=1.97$). It follows that at the spin flip time of $T_\pi = \SI{300}{\ns}$ that we use for our experiments, no scattering is expected to occur during the TOF at any interaction strength setting. The mean kinetic energies $E_\text{kin}$ (blue squares) show a similar dependence on $T_{\pi}$. This indicates again that only for $T_{\pi} \ll 1 / \lambda_\text{sc}$ (with $\lambda_\text{sc}\lesssim \SI{50}{\kilo\hertz}$) the true in-situ momentum distribution is obtained after the TOF sequence. The error bars are obtained from the counts in each bin of the momentum distribution and the correlation function respectively and by assuming poissonian statistics.}
    \label{fig:scat}
\end{figure}

To check that the switch into the almost non-interacting state is fast enough, we have studied the effect of its duration on the measured correlation signal (see Extended Data Fig.\,\ref{fig:scat}).
The measurement was taken at a fixed in-situ interaction strength $E_\text{B}/\hbar\omega_\text{r}=0.6$.
We find very good agreement of the decline in correlation strength with a model that assumes that each scattering event destroys all in-situ correlations of the two participating atoms.
The model contains no free parameters and depends just on the scattering rate $\lambda_\text{sc}$ of the $\ket{1}-\ket{3}$ mixture at the magnetic offset field ($B_\text{0}=\SI{750}{\gauss}$) and the in-situ density.
It enables us to extrapolate to measurements taken at the highest in-situ scattering rates (dashed line).
We conclude that for $T_\pi=\SI{300}{\ns}$ scattering during the TOF can be neglected for all interaction strength settings.

\begin{figure}
    \centering
	\includegraphics{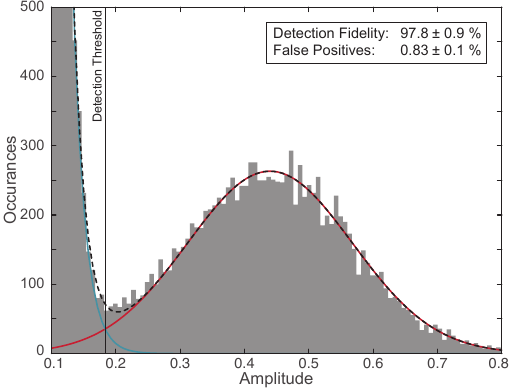}
    \caption{ \textbf{Single atom detection fidelity.}
    The raw images are analysed by first applying a low-pass filter followed by a simple peak detection algorithm.
    A histogram of the amplitudes of all detected peaks in 2000 images of a single spin component is plotted.
    We find a bimodal distribution.
    The maximum at low amplitudes originates from background noise of the camera.
    The second maximum at higher amplitudes is due to real photon clusters on the chip.
    Every peak with an amplitude above the threshold (vertical black line) is counted as an atom.
    This leads to single atom detection fidelities of $\SI{97.8\pm0.9}{\%}$.
    There is a probability of $\SI{5.0\pm0.5}{\%}$ for a false positive detection of an atom on each image for our chosen region of interested of $320\times 320\,\text{px}$.
    For 6+6 atoms this leads to a rate of false to true detections of $\SI{0.83\pm0.1}{\%}$.
    The solid red and blue lines are a Gaussian and exponential fits to the data respectively and the dashed line is their sum.
    }
    \label{fig:imFid}
\end{figure}

To image the two spin components of our gas on our camera we make use of the free-space imaging scheme discussed in detail in ref. \cite{Bergschneider_2018}.
It allows us to image single atoms in free space with high-fidelity and without any confining potential or cooling required.
Two counterpropagating illumination beams on the D2 line are used to excite the $^6$Li atoms and the emitted fluorescence light is collected through an objective ($NA=0.6$) on an EMCCD camera (see Fig.\,\ref{fig:mainSetup}c).
Each image is exposed for $\SI{15}{\us}$ and we collect around 20 photons per atom on the camera.
This leads to single atom detection fidelities on the order of $\SI{98}{\%}$ (see Extended Data Fig.\,\ref{fig:imFid}).
The different spin components are resolved by taking two images in quick succession (see Extended Data Fig.\,\ref{fig:imSeq}).
A sequence of radio frequency (RF) transitions is used to transfer atoms from each spin component to state $\ket{3}$ prior to their measurement.
The latter has a closed imaging transition and is therefore best suited for high-fidelity imaging.
When the first image is taken, the illumination beams are resonant only to atoms in state $\ket{3}$ and the other atoms in state $\ket{4}$ are so far detuned ($\sim \SI{2}{\giga\hertz}$) that off-resonant scattering is negligible.

\paragraph*{\textbf{Data analysis}}

\begin{figure*}
    \centering
	\includegraphics{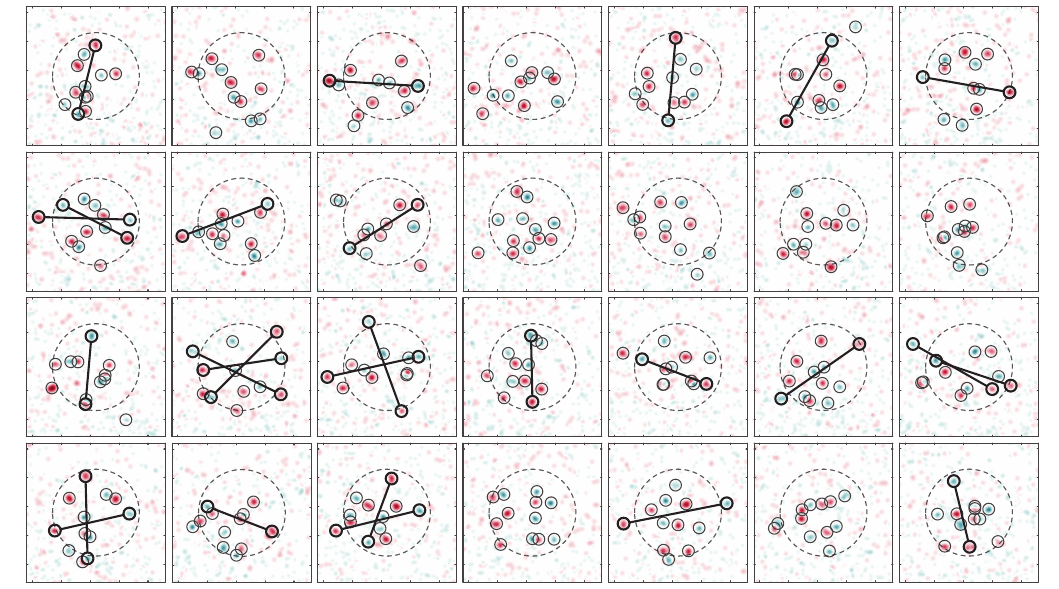}
    \caption{ \textbf{Collection of 28 single momentum space projections taken at $\bm{E_\text{B}/\hbar\omega_\text{r} = 1.97}$.} The images have been postselected for the correct particle number of the 6+6 ground state but are otherwise chosen randomly. The dashed circles indicate the Fermi momentum. Atoms pairs with $\Delta \phi < \ang{30}$ and $p_\uparrow$ and $p_\downarrow$ larger than $2/3p_\text{F}$ are highlighted. These are the particles that contribute to the pair correlation peak at the Fermi surface (see Fig.\,\ref{fig:mainC2}i). We find significantly more of such pairs in images taken at larger interaction strengths (compare Extended Data Fig.\,\ref{fig:pairsFew}).}
    \label{fig:pairsMany}
\end{figure*}

\begin{figure*}
    \centering
	\includegraphics{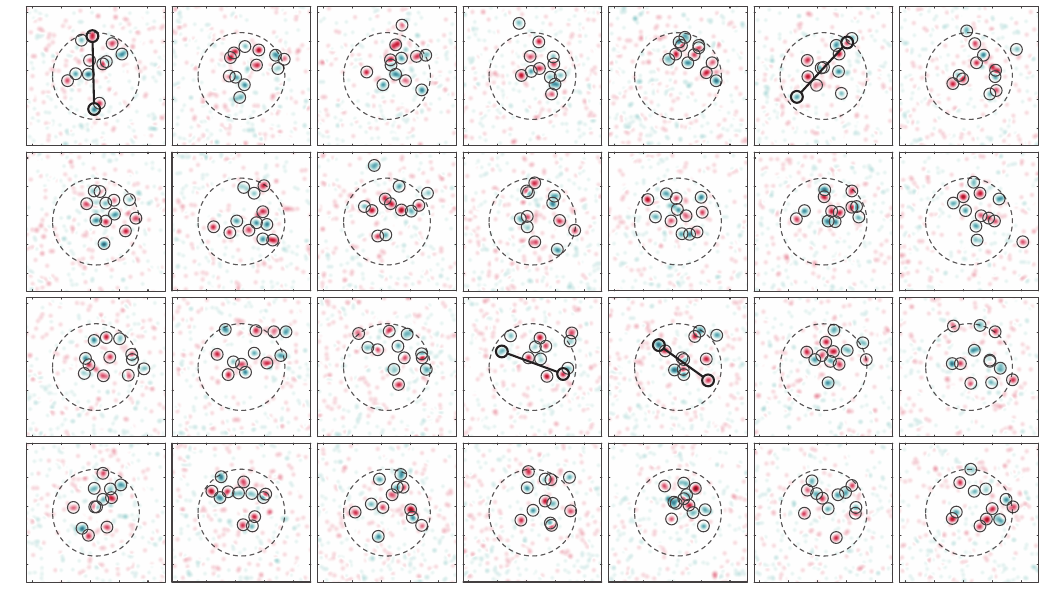}
    \caption{ \textbf{Collection of 28 single momentum space projections taken at $\bm{E_\text{B}/\hbar\omega_\text{r} = 0}$.} The images have been postselected for the correct particle number of the 6+6 ground state but are otherwise chosen randomly. The dashed circles indicate the Fermi momentum. All detected atom pairs with $\Delta \phi < \ang{30}$ and both $p_\uparrow$ and $p_\downarrow$ larger than $2/3p_\text{F}$ are highlighted. These are the particles that would contribute to the pair correlation peak at the Fermi surface (see Fig.\,\ref{fig:mainC2}f). Without interactions we find no additional pairs other than what is expected already from the single particle densities. Significantly more pairs are present in images taken at larger interaction strengths (compare Extended Data Fig.\,\ref{fig:pairsMany}).}
    \label{fig:pairsFew}
\end{figure*}

\begin{figure}
    \centering
	\includegraphics{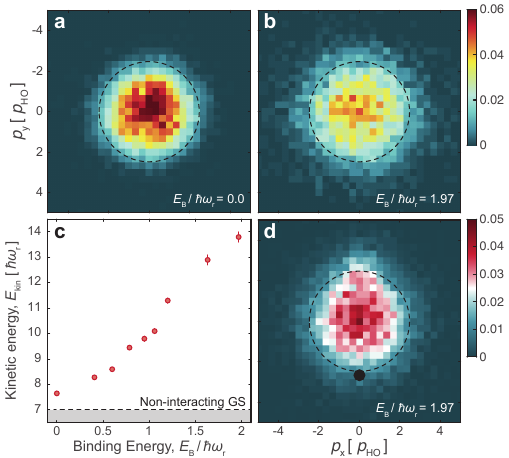}
    \caption{ \textbf{Average momentum space distributions.} \textbf{a,b,} The mean momentum space distributions $\braket{n_\uparrow(\bm{p}_\uparrow)}$ of a single spin component and averaged over 1000 images for two different binding energies and 6+6 atoms are shown. The dashed circle indicates the Fermi momentum. The distributions are to a good approximation radially symmetric. With increasing binding energy, the average momentum increases and we find more particles outside the Fermi momentum. This agrees with the picture that increasing the attraction allows particles to overcome the single particle gap and form first Cooper pairs that finally turn into tightly bound dimers. \textbf{c,} From the average momentum space distributions of both spin components it is straightforward to calculate the total mean kinetic energy of the system per spin component. For 6+6 non-interaction particles, we find a value very close to the expected ground state kinetic energy per spin component of $E^\text{gs}_\text{kin}=7\,\hbar\omega_\text{r}$. The kinetic energy increases monotonously as the attraction strength increases. The error bars represent the standard error of the mean. \textbf{d,} The unnormalized correlator $\mathcal{C}_{\overline{p}_\downarrow}^{(2)\,\ast}$ is shown for a $E_\text{B}/\hbar\omega_\text{r} = 1.97$. It defined as in equations \ref{eqn:c2} and \ref{eqn:c2_int} of the main text but without the $\braket{n}\braket{n}$ term. It shows that for strong enough binding energies the paired fraction becomes large enough that pair correlations are visible even without subtracting the single particle density contributions.}
    \label{fig:dens}
\end{figure}

From the EMCCD camera we obtain binary images with bright pixels where one or more photons hit the camera chip.
To analyse the images we first apply a low pass filter (see Extended Data Fig.\,\ref{fig:imSeq}c).
We search for peaks in these images and count all peaks above an optimized amplitude threshold as atoms (see Extended Data Fig.\,\ref{fig:imFid}).
The position of each atom in pixels is finally mapped to its in-situ momentum (see next section).
This results in a list of all momenta $\bm{p}_{\uparrow,i,j}$ and $\bm{p}_{\downarrow,i,j}$ of all atoms $i=1,...,6$ in each experimental run $j=1,..., N_\text{exp}\gtrsim 1000$ and for different interaction strength $E_\text{B}$ (see Extended Data Figs. \ref{fig:pairsMany} and \ref{fig:pairsFew}).
From this data we can directly obtain quantities like the average density in momentum space or the kinetic energy of the sample (see Extended Data Fig. \ref{fig:dens}).

To extract the correlation function $\mathcal{C}^{(2)}$, we transform to polar coordinates $\bm{p} \rightarrow (p,\phi)$ such that $\mathcal{C}^{(2)}(\bm{p}_\uparrow,\bm{p}_\downarrow) \rightarrow \mathcal{C}^{(2)}(p_\uparrow,\phi_\uparrow,p_\downarrow,\phi_\downarrow)$.
Due to the radial symmetry of our system $\mathcal{C}^{(2)}(p_\uparrow,\phi_\uparrow,p_\downarrow,\phi_\downarrow) \equiv \mathcal{C}^{(2)}(p_\uparrow,p_\downarrow,\Delta\phi)$ depends only on the difference between the angles of both particles $\Delta \phi = \phi_\uparrow - \phi_\downarrow$ but not the absolute values of $\phi_\uparrow $ and $\phi_\downarrow$.
We make use of this symmetry and integrate over all measurements with the same $\Delta \phi$ to increase the signal to noise ratio.
Finally, we bin the data according to the momentum of the ``reference'' particle $p_\downarrow \rightarrow \overline{p}_\downarrow \in \left[p_1,p_2\right]$ in momentum bins with a width of $\Delta p=1\,p_\text{HO}$.
Different choices for the momentum of the reference particle $\left[p_1, p_2\right]$ correspond to different 2D slices through the 4D correlation function $\mathcal{C}^{(2)}$ (compare Fig.\,\ref{fig:mainC2} a-e and f-j).
The slices can be expressed as $\mathcal{C}^{(2)}(p_\uparrow,\overline{p}_\downarrow,\Delta\phi) \equiv \mathcal{C}_{\overline{p}_\downarrow}^{(2)}(p_\uparrow,\Delta \phi )$ with
\begin{equation}
		\begin{aligned}
    &\mathcal{C}_{\overline{p}_\downarrow}^{(2)}(p_\uparrow,\Delta \phi )= \int_{p_1}^{p_2}\int_{0}^{2\pi}\int_{0}^{2\pi} \diff p'_\downarrow \diff \phi'_\downarrow \diff \phi'_\uparrow\\ &\quad\quad \mathcal{C}^{(2)}(p_\uparrow,\phi'_\uparrow,p'_\downarrow,\phi'_\downarrow)\,\delta(\Delta \phi - (\phi'_\uparrow-\phi'_\downarrow)) \, p'_\downarrow,
    \end{aligned}
\end{equation}
and are shown in Fig.\,\ref{fig:mainC2} of the main text.

\begin{figure*}
    \centering
	\includegraphics{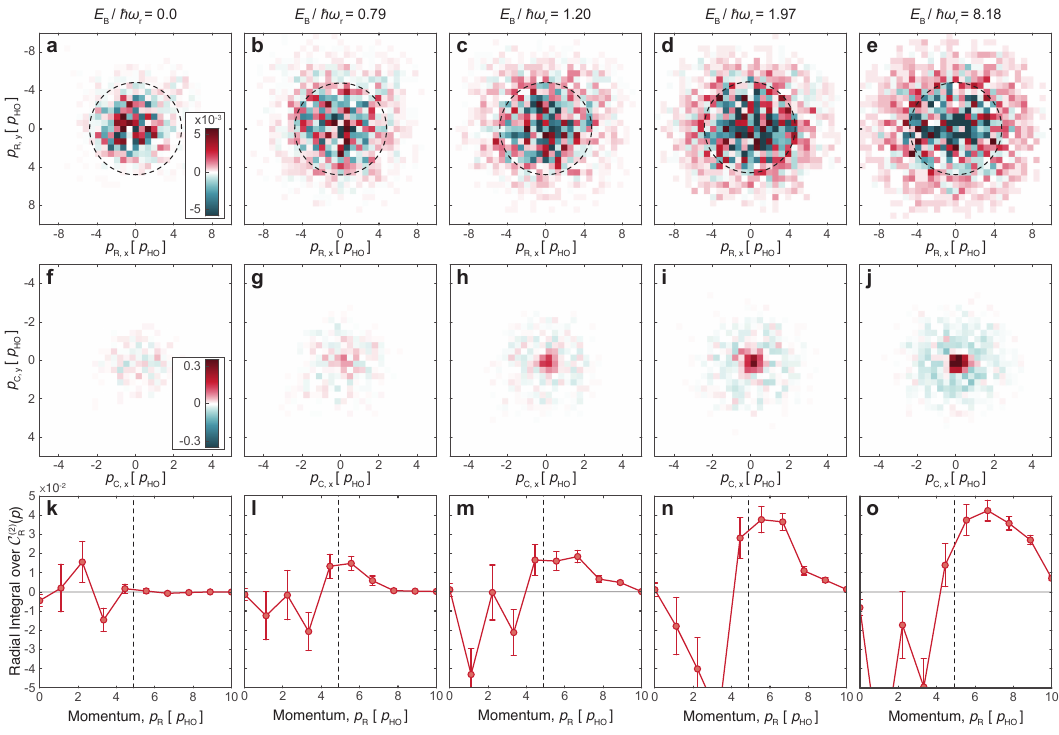}
    \caption{ \textbf{Alternative visualization of the density-density correlation function $\mathcal{C}^{(2)}$.}
    The pair correlation functions in relative ($\mathcal{C}^{(2)}_\text{R}$) and center of mass ($\mathcal{C}^{(2)}_\text{C}$) coordinates as a function of the interaction strength $E_\text{B}$ are shown in panels (a-e) and (f-j) respectively.
    The figure represent an alternative method of binning and visualizing the 4D correlation function $\mathcal{C}^{(2)}$ but is otherwise equivalent to the data shown in Fig.\,\ref{fig:mainC2} in the main text.
    The dashed circle is indicating twice the Fermi momentum $2p_\text{F}$.
    In relative coordinates (a-e), we find a surplus of particles with momenta of $\left| p_\text{R} \right| \approx 2 p_\text{F}$ as expected for the Formation of Cooper pairs with atoms located at the opposite ends of the Fermi surface.
    In the center of mass frame (f-j), emergence of pairing is indicated by a sharp peak at zero momentum that increases in weight with the interaction strength.
    Panels (k-o) show the radial integrals of the relative momentum correlation densities in panels (a-e).
    The error bars represent the standard error of the mean.
    }
    \label{fig:secondC2}
\end{figure*}

An alternative representation of the correlation function is possible by transforming to relative $\bm{p}_\text{R}= \bm{p}_\uparrow - \bm{p}_\downarrow$ and center of mass $\bm{p}_\text{C}=(\bm{p}_\uparrow + \bm{p}_\downarrow)/2$ coordinates $\mathcal{C}^{(2)}(\bm{p}_\uparrow,\bm{p}_\downarrow) \rightarrow \mathcal{C}^{(2)}(\bm{p}_\text{R},\bm{p}_\text{C})$.
By integrating over either one of the coordinates, we obtain the pair correlation function expressed in the relative $\mathcal{C}^{(2)}_\text{R}(\bm{p}_\text{R})=\int \diff \bm{p}_\text{C}\, \mathcal{C}^{(2)}(\bm{p}_\text{R},\bm{p}_\text{C})$ or center of mass $\mathcal{C}^{(2)}_\text{C}(\bm{p}_\text{C})=\int \diff \bm{p}_\text{R}\, \mathcal{C}^{(2)}(\bm{p}_\text{R},\bm{p}_\text{C})$ momentum coordinate respectively (see Extended Data Fig.\,\ref{fig:secondC2}).
The result is completely equivalent to the data in Fig.\,\ref{fig:mainC2} in the main text.
In the relative coordinate system, Cooper pair correlations appear at $\left| p_\text{R} \right| = \left| \bm{p}_\uparrow - \bm{p}_\downarrow\right| \approx 2 p_{F}$ (see Extended Data Fig.\,\ref{fig:secondC2} a-e).
In the correlation function $\mathcal{C}^{(2)}_\text{R}$, the pairing signal is spread out over a much larger area and the function is therefore much more sensitive to noise.
In addition, it is not possible to detect pairing inside the Fermi surface as $E_\text{B}\rightarrow\infty$ (d,e).
The correlation function is sensitive only to the number of pairs present at a given relative momentum $\left| p_\text{R} \right|$ and not to their relative angle $\Delta \phi$.
The center of mass correlation function $\mathcal{C}^{(2)}_\text{C}$ reveals the emergence of pairing as $E_\text{B}\rightarrow \infty$ (f-j).
We find a sharp peak centered at a zero center of mass momentum of the pairs with a weight that increases with $E_\text{B}$ as in Fig.\,\ref{fig:mainPairs}.

For Fig.\,\ref{fig:mainC2}a-j we postselect for only those runs where all atoms were detected in both images.
In Fig.\,\ref{fig:mainC2}k-o and \ref{fig:mainPairs} we also show data where up to three of the atoms were missed.
This significantly reduces the statistical errors of our results (due to the increased number of images used) and we have checked that we find no qualitative difference in the results for any of our measurements compared to a strict postselection.
The postselection rate of $\sim \SI{5}{\%}$ is an order of magnitude smaller for 6+6 atoms than what would be expected for our ground state preparation fidelity of $\SI{78}{\%}$ and single atom detection fidelity of $\SI{98}{\%}$.
The largest limitation is currently the low fidelity of the initial Raman transfer from state $\ket{3}$ to $\ket{4}$ ($\sim\SI{10}{\%}$ for the full system with 6+6 atoms).
The reasons are technical issues in the experiment that we plan to address and remove in the future.
Since imaging and preparation are completely independent in our experiment, the low probability for a simultaneous spin flip of all atoms in a single run does not affect the physical interpretation of our data.
The small postselection fidelity just increases the required total run time of the experiment to record some target image number where all atoms are present.

\paragraph*{\textbf{Experimental Parameters}}
We calculate the binding energy $E_\text{B}$ using the exact analytical solution of the Schrödinger equation for two ultracold atoms in a harmonic trap with axially symmetric confinement given in \cite{Idziaszek_2006}.
$E_\text{B}$ is dependent on three parameters: the radial- ($\omega_\text{r}$) and axial - ($\omega_\text{z}$) trap frequencies and the 3D scattering length $a_\text{3D}$.
We determine the trap frequencies using the trap modulation sequence described in \cite{Bayha_2020}.
The scattering length $a_\text{3D}$ is tuned by changing the magnetic offset field $B$ relative to the $\ket{1}-\ket{3}$ Feshbach resonance at $B_0 = \SI{690}{\gauss}$ \cite{Zuern_2013}. 

The TOF expansion takes place in an attractive potential created by the combination of an optical (Gaussian) and magnetic (harmonic) trap.
Due to the small anharmonicity of the Gaussian potential, the position after TOF and the in-situ momentum are --- contrary to the expansion in a purely harmonic potential --- not exactly related by a single scaling factor independent of the final position.
Nonetheless, since our experiments are performed in the limit where the atom cloud after TOF is much larger then in-situ, there exist a unique map of the final position to the in-situ momentum.
During the expansion our two in-situ spin components occupy low- ($\ket{1}-\ket{3}$) and high- ($\ket{4}$) field seeking states respectively.
For each of the two cases we obtain the momentum map separately by numerically solving the classical equations of motion in the resulting overall expansion potential.
Here, the small difference in the expansion times ($T_\text{TOF}^\uparrow=\SI{9} {\ms}$ and $T_\text{TOF}^\downarrow=\SI{9.2}{\ms}$) due to the small delay of the two images are also taken into account (see Extended Data Fig.\,\ref{fig:imSeq}b).
We obtain the overall expansion potential for the high-field (low-field) seeking states by adding (subtracting) the contribution of the magnetic trap to (from) the optical potential.
At the magnetic field of $B_\text{TOF} = \SI{750}{\gauss}$ that we use for expansion, the trap frequencies are given by $\omega_\text{opt} = 2\pi \times \SI{19.1 \pm 0.1}{\hertz}$ and $\omega_\text{mag} = 2\pi \times \SI{12.6 \pm 0.1}{\hertz}$ respectively.
Together with the magnification of $m=7.4$ of our optical setup, the solutions of the equations of motion allow us create two maps (one for each spin component) from camera pixels to momentum space in natural units of the in-situ harmonic oscillator $p_\text{HO} = \sqrt{\hbar m \omega_{r}}$.

The Fermi energy of the closed shell configurations is given by $E_\text{F}=(n_\text{F}+1)\hbar\omega_\text{r}$.
Here, $n_\text{F}=\sqrt{2N+1/4}-3/2$ denotes the quantum number of the highest completely filled harmonic oscillator level and $N= 1,3,6,10,...$ denotes the single-spin atom number of the given closed shell configuration.
We define the Fermi momentum using the continuum equation as $p_\text{F}=\sqrt{2mE_\text{F}}$.
The momentum distribution of the states in the highest filled shell $n_\text{F}$ is very broad for small particles numbers.
As a result, the Fermi momentum $p_\text{F}$ is in contrast to the Fermi energy $E_\text{F}$, not uniquely defined for mesoscopic samples in the harmonic oscillator.
This explains the large width on the order of $p_\text{HO}$ of the Cooper pair correlations at the Fermi surface, already in the weakly interacting limit (see Fig. \ref{fig:mainC2} f-j).
The center of mass momentum of the pairs fluctuates on the order of $p_\text{HO}$, explaining why they are not always detected with exactly opposite momenta (see Extended Data Figs. \ref{fig:secondC2} f-j and \ref{fig:pairsMany}).
Our definition of $p_\text{F}$ ensures that the correct value is reached in the limit $N\rightarrow\infty$ in a homogeneous system or when a local density approximation becomes applicable.
The ambiguity in the definition of $p_\text{F}$ and the fluctuations of the pair center of mass momentum do not affect the interpretation of our measurements in the mesoscopic system.
As the particle number is increased, we expect that the relative momentum uncertainty reduces continuously until it vanishes in the thermodynamic limit and only pairs with zero center of mass momentum remain at zero temperature.

\begin{figure}
    \centering
	\includegraphics{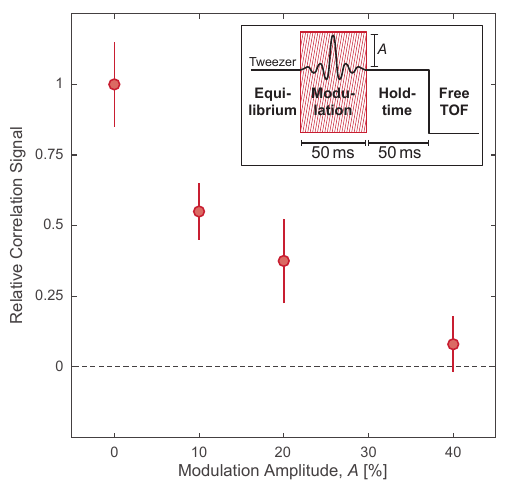}
    \caption{ \textbf{Correlations in a heated sample.} We increase the energy of the sample by modulating the radial confinement with a pulse of $\SI{50}{\ms}$ duration that is a square pulse of $\SI{700}{\hertz}$ width in frequency space and with variable amplitude $A$ (inset). We find that the pair correlations reduce with increasing energy of the sample until they vanish completely. This measurement was taken at intermediate binding energies of $E_\text{B}/\hbar\omega_\text{r} = 0.6$. In the future, we plan to study above ground state physics of our mesoscopic Fermi gas in more detail. To this end, we have to develop a precise method to measure temperatures of the sample. The error bars are obtained from the counts in each bin of the correlation function and by assuming poissonian statistics.}
    \label{fig:heat}
\end{figure}

We estimate that the temperature of our initial state is very low and the entropy per particle is on the order of $0.1\,k_\text{B}$.
When increasing the energy or temperature of the initial state by modulating the radial confinement, the amplitude of the pair correlation reduces significantly (see Extended Data Fig. \ref{fig:heat}).

\paragraph*{\textbf{BCS Theory}}
It is straightforward to calculate the density-density correlator $\mathcal{C}^{(2)}(\bm{p},-\bm{p})$ as defined by equation \ref{eqn:c2} in BCS theory. We recall the Bogoliubov transformations
\begin{align}
    c_{\bm{p}\uparrow}&=u_p\gamma_{\bm{p}\uparrow} - v_p\gamma^\dagger_{-\bm{p}\downarrow} \\
    c_{\bm{p}\downarrow}&=u_p\gamma_{\bm{p}\downarrow} + v_p\gamma^\dagger_{-\bm{p}\uparrow}
\end{align}
with $u^2_p=(1+\xi_p/E_p)/2$ and $v^2_p=(1-\xi_p/E_p)/2$. Here, $c^\dagger_{\bm{p},\text{s}}$ ($c_{\bm{p},\text{s}}$) is the fermionic creation (annihilation) operator of the particle with momentum $\bm{p}$ and spin s. The BCS spectrum is given by $E_p=\sqrt{\xi^2_p+\Delta^2}$ with $\xi_p=p^2/2m-E_\text{F}$ and for the pairing energy $\Delta$ we use the mean field value $\Delta = \sqrt{2E_\text{F}E_\text{B}}$ \cite{Randeria_1989}. Making use of the fact that the BCS ground state is free of excitations $\gamma_{\bm{p}} \ket{\Psi_{\text{BCS}}}=0$, we arrive at
\begin{equation}
	\begin{aligned}
    \mathcal{C}^{(2)}(\bm{p},-\bm{p}) &= \braket{c^\dagger_{\bm{p}\uparrow}c_{\bm{p}\uparrow} c^\dagger_{\bm{-p}\downarrow} c_{\bm{-p}\downarrow}} - \braket{c^\dagger_{\bm{p}\uparrow}c_{\bm{p}\uparrow} }\braket{c^\dagger_{\bm{-p}\downarrow} c_{\bm{-p}\downarrow}}\\ &= \mathcal{N}^2\frac{\Delta^2}{4(\xi_p^2+\Delta^2)}.
    \end{aligned}
\end{equation}
Here, the normalization factor $\mathcal{N}$ is chosen such that we obtain the correct total particle number at zero interactions ($\Delta =0$): $N_\uparrow=\int\braket{c^\dagger_{\bm{p}\uparrow}c_{\bm{p}\uparrow} }\diff\bm{p} =2\pi \mathcal{N} \int_0^\infty  v^2_p p \diff p$.

\paragraph*{\textbf{2D Molecules}}
A simple model for our system in the regime of strongest interactions ($E_\text{B}\gg E_\text{F}$) is to assume that all the particles form bosonic dimers that occupy the n=0 ground state of the 2D harmonic potential.
Following \cite{Zwierlein_2007}, an ansatz for the two body molecular wavefunction outside the scattering potential and in relative coordinates is
\begin{equation}
    \Psi_{\text{rel}}(r) = 
    \begin{cases}
        a_1 \times e^{-r/r_\text{B}}, & \text{for } r > r_B \\
        -\log{\frac{r}{r_\text{B}}} + a_2, & \text{for } r_{\text{0}} \leq r \leq r_\text{B} \\
        a_3 \times e^{-r^2 / r_b^2} + a_4, & \text{for } r < r_{\text{0}}
    \end{cases}
    \text{.}
\end{equation}

We choose the constants $a_{i}$ such that the wavefunction and its first derivative are continuous and properly normalized.
Here, $r$ is the interparticle distance and $r_\text{B}$ is the molecular binding length defined as $r_\text{B} = \hbar / \sqrt{2mE_\text{B}}$ with the atomic mass $m$ and the two-body binding energy $E_\text{B}$. 
We have inserted a short-distance cutoff $r < r_{\text{0}} = \SI{0.1}{r_\text{B}}$ to regularize the divergence of the logarithmic part for $r\rightarrow 0$.
We have checked that this cutoff does not affect the calculated pair correlations at small momenta.
The wavefunction for a single molecule is the product of the relative wavefunction $\Psi_{\text{rel}}$ and the center-of-mass wavefunction $\Psi_{\text{com}}$.
$\Psi_{\text{com}}$ is given by the two-dimensional ground state wavefunction of two particles in an harmonic oscillator.
The total wavefunction $\Psi_{\text{total}}(x_1, x_2, y_1, y_2)$ thus depends on four variables, namely the coordinates of the two particles.
To calculate the pair correlation signal that we expect for this trial wavefunction, we first perform a numerical Fourier transform in four dimensions.
This allows us to directly calculate $\mathcal{C}^{(2)}(\bm{p},-\bm{p})$ for a single dimer as defined in equation \ref{eqn:c2} of the main text.
Our model assumes that the molecules are completely independent.
The total correlation function $\mathcal{C}^{(2)}(\bm{p},-\bm{p})$ for $N+N$ particles is then just given by the summing over the contributions of $N$ single molecules.

%%%%%%%%%%%%%%%%%%%%%%%%%%%%%%%%%%%%%%%%%%%%%%%%%%%%%%%%%%%%%%%%%%%%%%%%%%%%%%%%%
%					     Extended Data Figure Captions      		   			%
%%%%%%%%%%%%%%%%%%%%%%%%%%%%%%%%%%%%%%%%%%%%%%%%%%%%%%%%%%%%%%%%%%%%%%%%%%%%%%%%%

\end{document}